\documentclass[11pt,a4paper]{article}
\pdfoutput=1
\usepackage{jheppub}
\usepackage{amsmath}
\usepackage{caption}
\usepackage{subcaption}

\newcommand{\be}{\begin{equation}}

\newcommand{\ee}{\end{equation}}

\newcommand{\bea}{\begin{eqnarray}}
\newcommand{\eea}{\end{eqnarray}}

\begin{document}
\begin{titlepage}
\thispagestyle{empty}

\vspace{2cm}
\begin{center}
\font\titlerm=cmr10 scaled\magstep4
\font\titlei=cmmi10
scaled\magstep4 \font\titleis=cmmi7 scaled\magstep4 {\Large{\textbf{Thermal quench of a dynamical QCD model in an external electric field}
\\}}
\setcounter{footnote}{0}
\vspace{1.5cm} \noindent{{
S. Heshmatian$^{a}$\footnote{e-mail:heshmatian@bzte.ac.ir }, F. Ahmadi$^{a}$ and A. Trounev$^{b}$
}}\\
\vspace{0.8cm}

{\it $^{a}$ Department of Engineering Sciences and Physics, Buein Zahra Technical University, Buein Zahra, Qazvin, Iran\\}
{\it $^{b}$ Department of Computer Technology and Systems, Kuban State Agrarian University, Krasnodar, Russia\\}

\vspace*{.4cm}
\end{center}
\vskip 2em
\setcounter{footnote}{0}
\begin{abstract}

In this article, we investigate the thermal equilibration of the holographic QCD model dual to the Einstein-Maxwell-Dilaton (EMD) gravity in the presence of an external electric field. The model captures the QCD features at finite temperature and finite chemical potential in both confinement and deconfinement phases and could be considered a good candidate to study the dynamics of the strongly interacting system in out-of-equilibrium conditions. For this purpose, we examine the instability imposed by an external electric field using the AdS/CFT dictionary and study the electric current flow and its relaxation for this holographic model. We study the effects of temperature, electric field strength, and chemical potential on the current flow of the stationary state by applying a constant electric field. Additionally, for a time-dependent electric field, we investigate the relaxation time scales of the system using equilibration time. Finally, we compare our results with those from other holographic models and experiments.

\end{abstract}
\end{titlepage}
\tableofcontents

\section{Introduction}
\label{sec1}

Gauge/gravity duality serves as a powerful tool for exploring out-of-equilibrium strongly coupled systems. The original duality establishes a correspondence between the ${\mathcal{N}}=4$ super Yang-Mills theory and Type IIB string theory in $AdS_5\times S^5$, and serves as a powerful tool to investigate the gauge theory at large Nc and strong coupling \cite{Maldacena:1997re}-\cite{CasalderreySolana:2011us}. However, given that the physical quantities in Quantum Chromodynamics (QCD) are temperature-dependent, significant efforts have been undertaken to extend this duality and develop more realistic holographic models capable of representing real-world QCD phenomena \cite{Polchinski:2000uf}-\cite{Panero:2009tv}. As perturbative QCD is only valid in the weak coupling regime and lattice QCD techniques are suitable for equilibrium conditions, the adoption of holographic QCD models has become prevalent to capture the behavior of strongly coupled dynamic systems.

Investigating the instabilities of the quark-gluon plasma (QGP) under the external electric fields is an interesting research subject, as experiments suggest the presence of strong electromagnetic fields after the QGP formation, which might influence the medium around the phase transition temperature \cite{Skokov:2009qp},\cite{Voronyuk:2011jd}. These strong electric fields could potentially lead to the creation of quark-antiquark pairs, analogous to the QED Schwinger effect where electron-positron pairs are created via the vacuum decay in the presence of an external electric field \cite{Schwinger:1951nm},\cite{Affleck:1981bma}. In the AdS/CFT context, this effect was initially explored by considering a rectangular or circular Wilson loop and calculating the quark-antiquark potential \cite{Semenoff:2011ng} and subsequently in \cite{Sato:2013iua}-\cite{Kawai:2015mha}. An alternative approach is the Hashimoto-Oka method \cite{Hashimoto:2013mua} which associates vacuum instability with the imaginary part of the effective action for the flavor probe brane.

The time evolution of a strongly coupled gauge theory system at a finite temperature is an interesting dynamical process investigated using the AdS/CFT correspondence. The system evolves from an initial Hamiltonian $H_0$ to the final state of a modified Hamiltonian due to the energy injection in a time interval. As the initial state is at non-zero temperature, this process is called the \textit{thermal quench}. This is a dynamic time-dependent process, and the system evolution is controlled by a characteristic time scale determining the energy injection rate \cite{Buchel:2012gw}-\cite{Caceres:2014pda}. If the energy is injected via an external electric field, the gauge field is mapped to an electric current in the boundary theory \cite{Hashimoto:2013mua}-\cite{Karch:2007pd} and applying a time-varying electric field, gives rise to a time-dependent current flow, which eventually settles into its steady state \cite{Hashimoto:2013mua}-\cite{Ebrahim:2015fku}.

Also, many attempts have been made to study various time scales of the quark-gluon plasma in the context of the AdS/CFT correspondence. For example, the isotropization time in far-from-equilibrium non-isotropic plasma has been studied using a time-dependent shear in the geometry and found the isotropization time to be of order $0.5~\text{fm/c}$ in Ref.~\cite{Chesler:2008hg}.

On the other hand, the confinement-deconfinement phase transition is an intriguing phenomenon of QCD that leads to the dissociation of quarkonium bound states near the phase transition temperature, as confirmed by the experimental findings \cite{PHENIX:2006gsi},\cite{ALICE:2013osk}. This suppression of quarkonium is more prominent at low energy densities, with the confined phase residing at lower temperatures and densities while the deconfined phase exists at higher temperatures and densities. Extensive theoretical studies have been performed to describe this phenomenon. Notably, lattice QCD calculations of the entropy of quark-antiquark pairs exhibit a peak near the phase transition region, indicating strong interactions \cite{Kaczmarek:2005zp}. To model the QCD confinement-deconfinement phase transition, a holographic QCD model has been proposed in \cite{Dudal:2017max} based on the Einstein-Maxwell-dilaton gravity model. This model provides a realistic framework for the QCD confinement-deconfinement phase transition with thermodynamic properties consistent with QCD and lattice results. The gravitational solution is at a finite temperature and indicates a first-order Hawking/Page phase transition from thermal-AdS to a black hole on the gravity side, corresponding to the standard confinement/deconfinement phases in the dual QCD theory. Furthermore, the model includes chemical potential, as the QCD confinement-deconfinement critical temperature and equation of state are sensitive to the chemical potential.

Motivated by the above contents, we are going to study the thermal quench of the holographic model \cite{Dudal:2017max} in the presence of external electric fields. For this purpose, we employ the case of standard confinement-deconfinement phase transition of the holographic model and consider two cases of constant and time-dependent applied electric fields. In the case of a constant external electric field, we calculate the stationary current in terms of temperature and electric field strength for different charge densities and chemical potential values. For the time-dependent case, the electric field is considered such that it starts from zero and reaches its final value. In this case, we determine the equilibration time for the thermal quench which is defined as the time when the time-dependent current approaches its final stationary value with $5\%$ uncertainty and stays in this regime afterward. The equilibration time dependency on the electric field and chemical potential is calculated numerically for a constant quench speed. Additionally, we examine the system's response in both fast and slow quench regimes, specified by small and large values of the characteristic time scale. For the fast quench regime, which corresponds to the small characteristic time scales, the universal behavior of the thermal quench is demonstrated. The rescaled equilibration time indicates an adiabatic behavior for the slow quench regime corresponding to the large characteristic time scales. 

The organization of the paper is as follows: In section~\ref{sec2}, a brief review of the standard confinement/deconfinement phase transition of the Einstein-Maxwell-dilaton model of \cite{Dudal:2017max} is presented. In section~\ref{sec3}, the stationary current flow of this dynamical holographic QCD model in the presence of a constant external electric field is studied in terms of the electric field strength and temperature for different chemical potential and charge density values. In section~\ref{sec4}, the response of the holographic model to a time-dependent electric field is investigated in terms of the electric field strength, temperature, and chemical potential values. Also, this dynamical process is studied in the fast quench and slow quench regimes in terms of the characteristic time scale $k$. Finally, we summarize our results in section~\ref{sec5}.

\section{Einstein–Maxwell-dilaton model}
\label{sec2}


In this section, we briefly review the holographic QCD model of \cite{Dudal:2017max} constructed from the Einstein-Maxwell-dilaton gravity model.\\

The Einstein-Maxwell-dilaton action in five dimensions is,
\begin{eqnarray}
&&S_{EM} = -\frac{1}{16 \pi G_5} \int \mathrm{d^5}x \sqrt{-g} \ \ \bigl[R-\frac{f(\phi)}{4}F_{MN}F^{MN} -\frac{1}{2}\partial_{M}\phi \partial^{M}\phi -V(\phi)\bigr].
\label{actionEF}
\end{eqnarray}
In this equation, $V(\phi)$ is the dilaton field potential, $f(\phi)$ is a gauge kinetic function representing the coupling between dilaton and gauge field $A_{M}$, and $G_5$ is the Newton constant. From the above action, the Einstein, Maxwell, and dilaton equations of motion could be obtained as,
\begin{eqnarray}
R_{MN}-\frac{1}{2}g_{MN}R-T_{MN}=0,
\label{Einsteineq}
\end{eqnarray}
\begin{eqnarray}
\nabla_{M}[f(\phi) F^{MN}]=0,
\label{Maxwelleq}
\end{eqnarray}
\begin{eqnarray}
\partial_{M} \bigl[ \sqrt{-g}\partial^{M}\phi \bigr]-\sqrt{-g} \biggl( \frac{\partial V}{\partial \phi} + \frac{F^2}{4}\frac{\partial f}{\partial \phi} \biggr)=0
\label{dilatoneq}
\end{eqnarray}
where
$$T_{MN}=\frac{1}{2} \biggl(\partial_{M}\phi \partial_{M}\phi-\frac{1}{2}g_{MN} (\partial \phi)^2 -g_{MN}V(\phi) \biggr)+\frac{f(\phi)}{2}\biggl(F_{MP}F_{N}^{\ P} -\frac{1}{4}g_{MN}F^2 \biggr).$$
By considering the following ansatz for the metric, gauge field, and dilaton field,
\begin{eqnarray}
& & ds^2=\frac{L^2 e^{2 A(z)}}{z^2}\biggl(-g(z)dt^2 + \frac{dz^2}{g(z)} + dy_{1}^2+dy_{3}^2+dy_{3}^2 \biggr)\,, \nonumber \\
& & A_{M}=A_{t}(z), \ \ \ \ \phi=\phi(z),
\label{metricansatz}
\end{eqnarray}
the eqs.~(\ref{Einsteineq}), (\ref{Maxwelleq}) and (\ref{dilatoneq}) could be solved simultaneously to obtain the following equations,
\begin{eqnarray}
\phi'' +\phi' \left( -\frac{3}{z}+\frac{g'}{g}+3A' \right)- \frac{L^2 e^{2A}}{z^2 g} \frac{\partial V}{\partial \phi}+\frac{z^2 e^{-2A}A_{t}'^{2}}{2 L^2 g} \frac{\partial f}{\partial \phi}=0,
\label{phieq}
\end{eqnarray}
\begin{eqnarray}
A_{t}'' +A_{t}' \left( -\frac{1}{z}+\frac{f'}{f}+A' \right)=0,
\label{Ateq}
\end{eqnarray}
\begin{eqnarray}
g'' + g'\biggl ( -\frac{3}{z} + 3A' \biggr)- \frac{e^{-2A} A_{t}'^{2} z^2 f}{L^2}=0,
\label{geq}
\end{eqnarray}
\begin{eqnarray}
A''+\frac{g''}{6g} + A' \left(-\frac{6}{z}+\frac{3g'}{2g}\right)-\frac{1}{z}\left(-\frac{4}{z}+\frac{3g'}{2g}\right)
+3 A'^{2} + \frac{L^2 e^{2A} V}{3z^2 g}=0,
\label{Aeq}
\end{eqnarray}
\begin{eqnarray}
A''- A'\left(-\frac{2}{z}+A'\right)+\frac{\phi'^2}{6}=0
\label{Aeq2}
\end{eqnarray}
Here, it is assumed that various fields depend only on the extra radial coordinate $z$. Also, $z=0$ corresponds to the asymptotic boundary of the spacetime and $L$ is the AdS length scale. Out of these differential equations, it is consistent to consider only four equations and the eq.~(\ref{phieq}) is considered as a constraint equation. Subsequently, other equations could be solved analytically in terms of functions $A(z)$ and $f(z)$ as,
\begin{eqnarray}
&&g(z)=1-\frac{\int_{0}^{z} dx \ x^3 e^{-3A(x)} \int_{x_c}^{x} dx_1 \ \frac{x_1 e^{-A(x_1)}}{f(x_1)}}{\int_{0}^{z_h} dx \ x^3 e^{-3A(x)} \int_{x_c}^{x} dx_1 \ \frac{x_1 e^{-A(x_1)}}{f(x_1)} }, \nonumber \\
&&\phi'(z)=\sqrt{6(A'^2-A''-2 A'/z)}, \nonumber \\
&& A_{t}(z)=\sqrt{\frac{-1}{\int_{0}^{z_h} dx \ x^3 e^{-3A(x)} \int_{x_c}^{x} dx_1 \ \frac{x_1 e^{-A(x_1)}}{f(x_1)}}} \int_{z_h}^{z} dx \ \frac{x e^{-A(x)}}{f(x)}, \nonumber \\
&&V(z)=-\frac{3z^2ge^{-2A}}{L^2}\bigl[A''+A' \bigl(3A'-\frac{6}{z}+\frac{3g'}{2g}\bigr)-\frac{1}{z}\bigl(-\frac{4}{z}+\frac{3g'}{2g}\bigr)+\frac{g''}{6g} \bigr]
\label{metricsol}
\end{eqnarray}
Here, the boundary condition is considered such that $g(z_h)=0$ at the horizon and $g(z)\to1$ at the asymptotic boundary. The integration constant $x_c$ in eq.~(\ref{metricsol}) is fixed in terms of the chemical potential of the boundary theory. Chemical potential could be obtained by expanding $A_{t}$ near the asymptotic boundary $z=0$ and using the gauge/gravity mapping as,
\begin{eqnarray}
\mu=-\sqrt{\frac{-1}{\int_{0}^{z_h} dx \ x^3 e^{-3A(x)} \int_{x_c}^{x} dx_1 \ \frac{x_1 e^{-A(x_1)}}{f(x_1)}}} \int_{0}^{z_h} dx \ \frac{x e^{-A(x)}}{f(x)}.
\label{mueq}
\end{eqnarray}
In order to maintain the linear Regge trajectories for the discrete spectrum of the mesons in the boundary theory, the arbitrary function $f(z)$ is considered as the following form,
\begin{eqnarray}
f(z)=e^{-c z^2 -A(z)},
\label{fansatz}
\end{eqnarray}
Here, the constant $c=1.16~{\text GeV}^2$ is fixed by matching the holographic mass spectrum of heavy meson bound states to that of the lowest lying heavy meson states \cite{He:2013qq}. Using eq.~(\ref{fansatz}) in eq.~(\ref{metricsol}), the gravity solution is given by
\begin{eqnarray}
&&g(z)=1-\frac{1}{\int_{0}^{z_h} dx \ x^3 e^{-3A(x)}} \biggl[\int_{0}^{z} dx \ x^3 e^{-3A(x)} + \frac{2 c \mu^2}{(1-e^{-c z_{h}^2})^2} \det \mathcal{G} \biggr],\nonumber \\
&&\phi'(z)=\sqrt{6(A'^2-A''-2 A'/z)}, \nonumber \\
&& A_{t}(z)=\mu \frac{e^{-c z^2}-e^{-c z_{h}^2}}{1-e^{-c z_{h}^2}}, \nonumber \\
&&V(z)=-\frac{3z^2ge^{-2A}}{L^2}\left[A''+A' \left(3A'-\frac{6}{z}+\frac{3g'}{2g}\right)-\frac{1}{z}\left(-\frac{4}{z}+\frac{3g'}{2g}\right)+\frac{g''}{6g} \right]
\label{metricsol1}
\end{eqnarray}
where
\[
\det \mathcal{G} =
\begin{vmatrix}
\int_{0}^{z_h} dx \ x^3 e^{-3A(x)} & \int_{0}^{z_h} dx \ x^3 e^{-3A(x)- c x^2} \\
\int_{z_h}^{z} dx \ x^3 e^{-3A(x)} & \int_{z_h}^{z} dx \ x^3 e^{-3A(x)- c x^2}
\end{vmatrix}.
\]
The Hawking temperature and entropy of the black hole solution of eq.~(\ref{metricsol1}) are given by,
\begin{eqnarray}
T&=& \frac{z_{h}^3 e^{-3 A(z_h)}}{4 \pi \int_{0}^{z_h} dx \ x^3 e^{-3A(x)}} \biggl[ 1+\frac{2 c \mu^2 \bigl(e^{-c z_h^{2}}\int_{0}^{z_h} dx \ x^3 e^{-3A(x)}-\int_{0}^{z_h} dx \ x^3 e^{-3A(x)}e^{-c x^{2}} \bigr)}{(1-e^{-c z_h^{2}})^2} \biggr] \nonumber \\
S&=& \frac{L^3 e^{3 A(z_h)}}{4 G_5 z_{h}^3}.
\label{Htemp}
\end{eqnarray}

\begin{figure}
\begin{subfigure}{.5\textwidth}
\centering
\includegraphics[width=3in,height=2.3in]{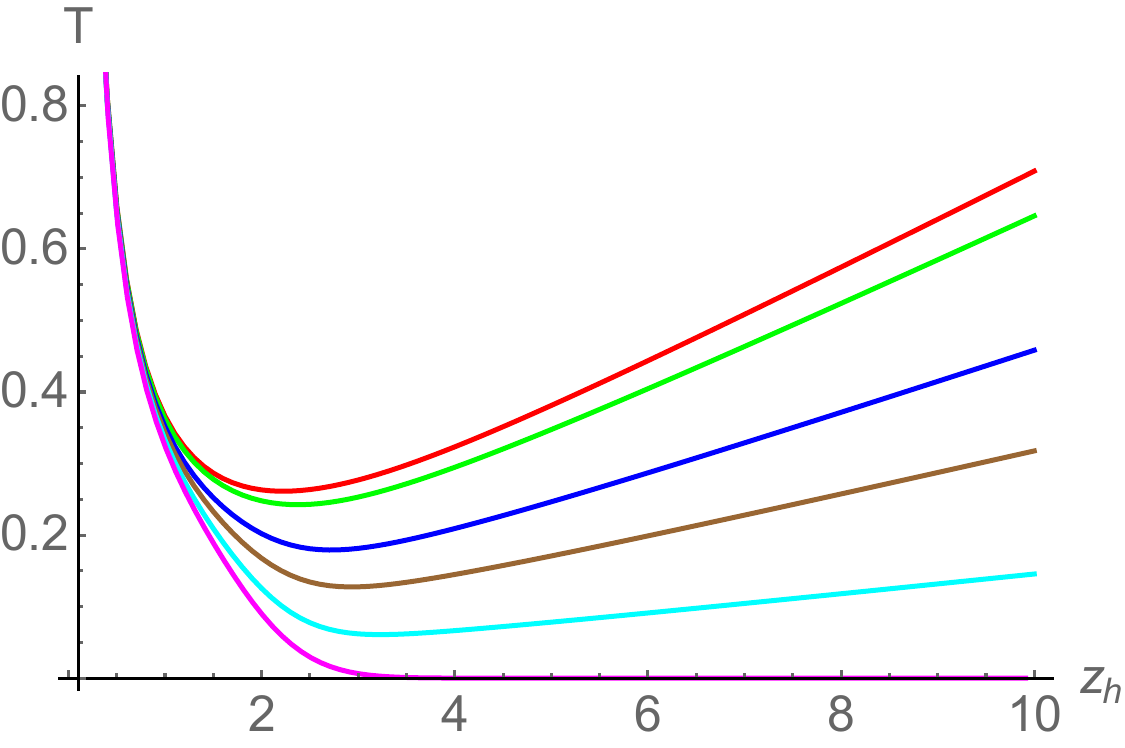}
\caption{}
\label{zhvsTblackhole}
\end{subfigure}
\begin{subfigure}{.5\textwidth}
\centering
\includegraphics[width=3in,height=2.3in]{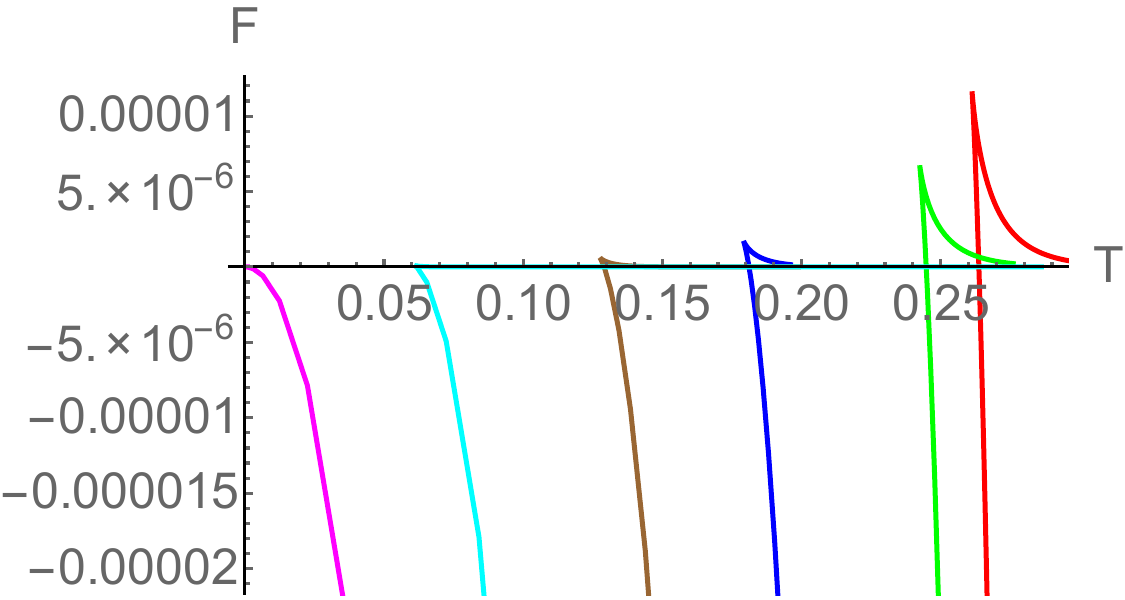}
\caption{}
\label{TvsFblackhole}
\end{subfigure}
\caption {\small (\subref{zhvsTblackhole}) Hawking temperature $T$ as a function of $z_h$ for various values of the chemical potential $\mu$. (\subref{TvsFblackhole}) Normalized free energy $F$ as a function of temperature for various values of the chemical potential $\mu$. Red, green, blue, brown, and cyan curves correspond to $\mu=0$, $0.2$, $0.4$, $0.5$, $0.6$, and $0.673$ respectively \cite{Dudal:2017max}.}
\label{zTF}
\end{figure}

In holographic QCD models, the appropriate form of the scale factor $A(z)$ is determined to be compatible with certain QCD features like linear Regge trajectories and the QCD phase diagram. In \cite{Andreev:2006nw}-\cite{Arefeva:2018hyo}, polynomial functions have been chosen for heavy quarks and in \cite{Gursoy:2010fj}, \cite{Li:2017tdz} and \cite{Arefeva:2020byn}, logarithmic functions of the radial coordinate have been suggested for light quarks. Depending on the arbitrary function $A(z)$, different kinds of phase transitions occur on the gravity side corresponding to different phase transitions in the boundary theories. In \cite{Dudal:2017max}, a quadratic function for $A(z)$ has been chosen as,
\begin{eqnarray}
A(z)=- \bar{a} z^2\, ,
\label{AansatzA2}
\end{eqnarray}
for which, there exists a first-order Hawking/Page phase transition from thermal-AdS to black hole which corresponds to the standard confinement/deconfinement phase transition in the dual QCD theory. The parameter $\bar{a}=c/8 \simeq 0.145$ has been fixed such that the critical temperature of the phase transition is around $0.27~\text{GeV}$ at zero chemical potential. By this choice, the analytic solution could be obtained from eq.~(\ref{metricsol1}) for the dilaton field as,
\begin{eqnarray}
\phi(z)=z \sqrt{3 \bar{a}(3+2 \bar{a}z^2)} +3 \sqrt{\frac{3}{2}}\sinh ^{-1} \biggl[ {\sqrt{\frac{2 \bar{a}}{3}}z} \biggr],
\label{phisol}
\end{eqnarray}
and the eq.~(\ref{fansatz}) gets the following form,
\begin{eqnarray}
f(z)=e^{-(c+\bar{a}) z^2}.
\label{fsol}
\end{eqnarray}
One could also write the near boundary expansion for the dilaton field and the dilaton potential as follows,
\begin{eqnarray}
&&\phi(z)=6 \sqrt{\bar{a}}z +2/3 \bar{a}^{3/2} z^3+\ldots, \nonumber \\
&&V(\phi)=-\frac{12}{L^2}+\frac{\Delta(\Delta-4)}{2}\phi^2(z)+\ldots, \ \ \Delta=3 \, ,
\label{phiV-nb}
\end{eqnarray}
where the dimension of the dual operator is $\Delta=3$ and the dilaton mass is $m^2=\Delta(\Delta-4)$.\\
The Hawking temperature in terms of the horizon radius $z_h$ for various values of chemical potential is shown in Fig.~\ref{zhvsTblackhole}. From this figure, one could find that for the stable black hole phase (with small $z_h$), the slope is negative while, for the unstable thermal AdS phase (with large $z_h$), the slope is positive. Below a critical chemical potential $\mu_c=0.673~\text{GeV}$, only the unstable branch exists. The stable black hole solution does not exist below a certain minimal temperature $T_{min}$ for each $\mu$ which suggests a phase transition from black hole to thermal-AdS as the Hawking temperature decreases. The free energy has been normalized such that the free energy of the thermal-AdS is zero and is shown in Fig.~\ref{TvsFblackhole} for different values of chemical potential. The phase transition appears in this diagram where the free energy is positive for the unstable branch and becomes negative after some critical temperature $T_{crit}$ along the stable branch. 

\section{Stationary current}
\label{sec3}
In this section, we study the current flow of the holographic QCD model of \cite{Dudal:2017max} in the presence of a constant external electric field using the Hashimoto-Oka method of \cite{Hashimoto:2013mua}.

In holography, for each source on the boundary theory, there exists a corresponding field on the gravity side. Accordingly, the local gauge field in the D7-brane action is dual to the global symmetry of the fermion number in the boundary theory. Therefore, applying an electric field leads to a current flow in the boundary theory as the quarks have an electric charge. In order to have $N_f$ quark flavors in the boundary theory, $N_f$ D7-branes are added to the gravitational background \cite{Karch:2002sh}. The D7-branes extend along $AdS_5\times S^3$ and wrap the $S^3$ inside a five-dimensional sphere $S^5$ whose metric is given by,
\begin{eqnarray}
d\Omega_5^2= L^2(d\theta^2+\cos^2\theta d\Omega_3^2+\sin^2\theta d\varphi^2).
\end{eqnarray}
Localization of the D7-brane is parametrized by $\phi=constant$ and the embedding function $\theta(z)$ which is dual to the hypermultiplet mass operator and runs from zero for a massless quark to $\pi/2$ for a quark with large mass.

Here, we are interested in the probe limit for simplicity. In this limit, a finite number of flavors exist in the gauge theory with a large number of color charges so that $N_f\ll N_c$. It means that the small number of probe D7-branes in the gravity side, does not change the gravitational background and one could neglect the back-reaction of the probe brane on the background geometry. This limit is known as the quenched approximation in the lattice gauge theory in which the back-reaction of quarks on the gluon vacuum is neglected and the Wess-Zumino couplings are assumed to be zero \cite{Karch:2002sh}. Therefore, the dynamics of the gauge fields living on the probe D7-brane are governed only by the DBI action as,
\begin{eqnarray}
S_{\rm D7} = - \mu_7 \int\! dt d^3\vec{x} dz d\Omega_3 \, \sqrt{-\det
\left[
g_{ab} + 2\pi \alpha' F_{ab}
\right]}\, ,
\label{DBI}
\end{eqnarray}
where $F_{ab}$ is the field strength of the gauge field living on the probe brane and $\mu_7$ is the brane tension. Also, $g_{ab}=G_{MN}\partial_a X^M \partial_b X^N$ is the induced metric on the brane where $a, b$ represent the D7-brane coordinates ($ t, \vec{x}, z, \Omega_3)$ and $M,N$ represent the background coordinates ($t, \vec{x}, z, \Omega_5$).

Using eq.~(\ref{DBI}) leads to the following effective Lagrangian in terms of the metric functions and field strengths,
\begin{eqnarray}
{\cal{L}}&\propto& \int dz \, (G_{xx}^{3} G_{\Omega\Omega}^{3} G_{tt} G_{zz})^{1/2} \sqrt{\chi} ,\nonumber \\
\chi &=& 1- \frac{(2\pi \alpha')^2}{G_{xx} G_{tt} G_{zz}} \biggl(G_{xx} F_{tz}^2 +G_{zz} F_{tx}^2-G_{tt} F_{zx}^2 \biggr ).
\label{Lag}
\end{eqnarray}
From the Lagrangian (\ref{Lag}), the following equations of motion for the gauge fields are obtained,
\begin{eqnarray}
\partial_z\left(\sqrt{\frac{ G_{xx}^3}{{ G_{tt} G_{zz}}}}\frac{ F_{tz}}{\sqrt{\chi}}\right)=0 \ \ , \ \ \ \
\partial_t\left(\sqrt{\frac{ G_{xx}^3}{{ G_{tt} G_{zz}}}}\frac{ F_{tz}}{\sqrt{\chi}}\right)&=&0 , \nonumber\\
\partial_z\left(\sqrt{\frac{ G_{xx} G_{tt} }{G_{zz}}}\frac{ F_{zx}}{\sqrt{\chi}}\right)-\partial_t\left(\sqrt{\frac{G_{xx} G_{zz} }{{ G_{tt} }}}\frac{ F_{tx}}{\sqrt{\chi}}\right)&=&0 ,
\label{eom1}
\end{eqnarray}
where, we have considered the embedding function $\theta(z)=0$ in order to a have massless hypermultiplets. Therefore, $ G_{\Omega\Omega}=L^2$ and we have the flat embedding which always solves the equation of motion derived from the DBI action. For a time-independent electric field, $\partial_t =0$ and according to the AdS/CFT dictionary, the resulting integration constants of the above equation, corresponding to the electric current flow and the charge density \cite{Karch:2007pd}-\cite{Erdmenger:2007bn}. Therefore, one can define the dimensionless current flow $j$ and charge density $d$ as,
\begin{eqnarray}
j= \sqrt{\frac{ G_{xx} G_{tt} }{G_{zz}}}\frac{2\pi \alpha' F_{zx}}{\sqrt{\chi}}\, , \ \
d= \sqrt{\frac{ G_{xx}^3}{{ G_{tt} G_{zz}}}}\frac{2\pi \alpha' F_{tz}}{\sqrt{\chi}}.
\label{jd}
\end{eqnarray}\\
Using these equations to write $F_{zx}$ and $F_{tz}$ in terms of $j$ and $d$ and inserting into the $\chi$ definition eq.~(\ref{Lag}) leads to,
\begin{eqnarray}
\chi = \frac{1-\frac {(2\pi \alpha')^2 E_0^2}{G_{tt}G_{xx}}}{1+\frac{1}{G_{xx}^3}\biggl(d^2 - j^2 G_{xx}G_{tt}^{-1} \biggr )} .
\label{chi}
\end{eqnarray}
where $E_0=F_{tx}$ is the constant external electric field. In order to find the stationary current flow, one can use the reality condition for the D-brane Lagrangian \cite{Karch:2007pd}-\cite{Erdmenger:2007bn}. Hence, one can write the following equations for the numerator and denominator of $\chi$ to change sign simultaneously in a specific radial coordinate $z=z_p$,
\begin{eqnarray}
\biggl[1-\frac {(2\pi \alpha')^2 E_0^2}{G_{tt}G_{xx}}\biggr]_{z=z_p}=0 ,
\label{zp1}
\end{eqnarray}\\
\begin{eqnarray}
\biggl[1+\frac{1}{ G_{xx}^3}\biggl(d^2 - j^2 G_{xx}G_{tt}^{-1} \biggr )\biggr]_{z=z_p}=0 .
\label{jd0}
\end{eqnarray}
From eq.~(\ref{jd0}), we find the following expression for the stationary current $j\equiv j_0$,
\begin{eqnarray}
j_0=\biggl[\sqrt{d^2 G_{tt}G_{xx}^{-1}+G_{tt} G_{xx}^2}\biggr]_{z=z_p} .
\label{js}
\end{eqnarray}

This non-zero current indicates a real effective action for the D7-brane and also an event horizon formation in $z=z_p$. In order to compute the holographic stationary current $j_0$, we have obtained $z_p$ numerically from eq.~(\ref{zp1}) and inserted it into eq.~(\ref{js}) for different values of the parameters. In our calculation, we have used eq.~(\ref{AansatzA2}) for the scale function which leads to the standard confinement/deconfinement phase transition in the holographic boundary of \cite{Dudal:2017max}. The AdS radius is set to 1, and $T$ is the temperature of the initial thermal state given by the Hawking temperature of the black hole solution eq.~(\ref{Htemp}).

\begin{figure}
\begin{subfigure}{.5\textwidth}
\centering
\includegraphics[width=3in,height=2.3in]{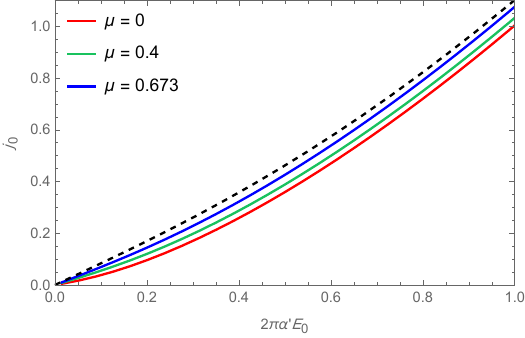}
\caption{}
\label{j0E}
\end{subfigure}
\begin{subfigure}{.5\textwidth}
\centering
\includegraphics[width=3in,height=2.3in]{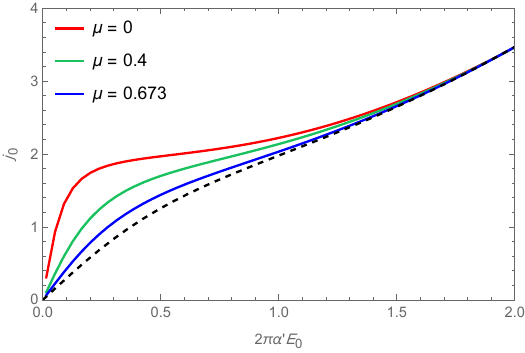}
\caption{}
\label{j0Ed}
\end{subfigure}
\caption {\small The dimensionless stationary current $j_0$ in terms of $2\pi \alpha'E_0$ for various values of the chemical potential $\mu$ at $T=0.27~\text{GeV}$. Red, green, and blue curves represent $\mu=0$, $0.4$, and $0.673$ respectively In units \text{GeV} and the dashed curves represent the $I-V$ curve for the Hashimoto-Oka model. (\subref{j0E}) non-doped case $d=0$ and (\subref{j0Ed}) doped case $d=2$.}
\label{j0Emu}
\end{figure}

The holographic current in terms of the rescaled external electric field, $2\pi \alpha'E_0$, is shown in Fig.~\ref{j0Emu}. This figure is depicted for non-dopped and doped systems for three different values of chemical potential $\mu$ at $T_c=0.27~\text{GeV}$ which is the critical temperature of the standard confinement/deconfinement phase transition at zero chemical potential and the minimum temperature for which there always exists the black hole solution for all $\mu$ on the gravity side. The dashed curves correspond to the stationary current $j_0$ in terms of $2\pi \alpha'E_0$ for the massless supersymmetric QCD model which has been called the $I-V$ curve in the Hashimoto-Oka model of ref.~\cite{Hashimoto:2013mua}. From these figures, we find that our results for the holographic model are less for the non-doped case with $d=0$ and more for the doped case with $d=2$ compared to the results from the Hashimoto-Oka model. Likewise, one can observe that in both cases, the curves for $\mu=0.673$ are closer to the stationary current curves of \cite{Hashimoto:2013mua} and for large values of the electric field, all currents converge to the $I-V$ curves of the Hashimoto-Oka model. \\
\begin{figure}
\begin{subfigure}{.5\textwidth}
\centering
\includegraphics[width=3in,height=2.3in]{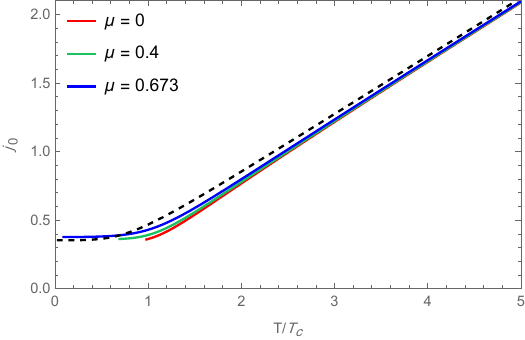}
\caption{}
\label{j0T}
\end{subfigure}
\begin{subfigure}{.5\textwidth}
\centering
\includegraphics[width=3in,height=2.3in]{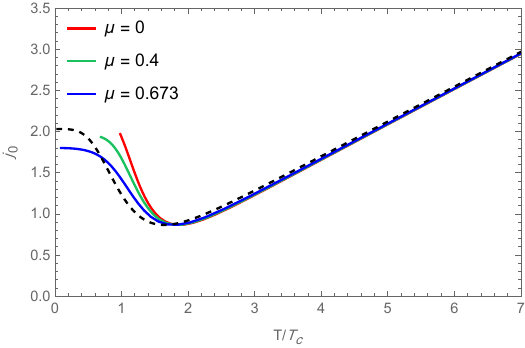}
\caption{}
\label{j0Td}
\end{subfigure}
\caption {\small The dimensionless stationary current $j_0$ versus $T$ for various values of the chemical potential $\mu$ at $E_0=0.5~\text{GeV}$. Red, green, and blue curves represent $\mu=0$, $0.4$, and $0.673$ respectively In units \text{GeV} and the dashed curves represent the $I-V$ curve results of the Hashimoto-Oka model. (\subref{j0T}) non-doped case $d=0$ and (\subref{j0Td}) doped case $d=2$.}
\label{j0Tmu}
\end{figure}
Also, the stationary current in terms of the rescaled temperature $T/T_c$ is depicted in Fig.~\ref{j0Tmu} for $E_0=0.5$ and three different values of chemical potential $\mu$. For each value of $\mu$, the corresponding curve starts from a critical temperature $T_{crit}$ which is determined by the free energy diagram of Fig.~\ref{TvsFblackhole}. A non-trivial behavior could be found at lower temperatures accessible for larger $\mu$, where the current is larger (less) than the Hashimoto-Oka model for $d=0$ ($d=2$) case. A similar behavior to the $j_0-E_0$ figures could be observed from these diagrams for $T \gtrsim 0.2$, and all currents converge to the Hashimoto-Oka model curve at high temperatures.
\section{Time dependent current}
\label{sec4}
In this section, we are going to investigate the dynamical response of the holographic QCD model to an external time-dependent electric field. In this phenomenon, the system experiences a time-dependent energy injection, which gives rise to a time-dependent current flow $j(t)$ in the boundary theory. The initial state is at a finite temperature and after a while, the system relaxes into the final steady state current $j_0$ corresponding to the final amount of the electric field. For this purpose, we use  the equilibration time as a time scales. 

The  equilibration time, $t_{eq}$, which is defined as the first time that the time-dependent response (which is the electric current in our case) reaches its final stationary value with $5\%$ uncertainty and remains within this regime afterward \cite{Buchel:2012gw}. Using this time scale enables us to find out how the energy injection rate affects the system evolution and relaxation as it depends on the quench speed. Here, we use the following time-dependent error parameter,
\begin{eqnarray}
\delta (t) =\frac{j(t) - j_{0}}{j_{0}},
\label{error}
\end{eqnarray}
therefore, $\delta (t_{eq}) < 0.05$ for all $t>t_{eq}$.

In order to calculate the equilibration time, we have to find the time-dependent current for different parameters. For this purpose, we use the Lagrangian (\ref{Lag}) to obtain the equations of motion for the gauge field $A_x$ as follows,
\begin{eqnarray}
\partial_z\left(\sqrt{\frac{ G_{xx} G_{tt} }{G_{zz}}}\frac{ F_{zx}}{\sqrt{\chi}}\right)-\partial_t\left(\sqrt{\frac{G_{xx} G_{zz} }{{ G_{tt} }}}\frac{ F_{tx}}{\sqrt{\chi}}\right)&=&0 ,\nonumber \\
\chi =1- \frac{(2\pi \alpha')^2}{G_{xx} G_{tt} G_{zz}} \biggl(G_{zz} F_{tx}^2-G_{tt} F_{zx}^2 \biggr ),
\label{eomax}
\end{eqnarray}
where we have used the gauge choice $A_z=0$ and assumed the charge density $d$ equal to zero. In order to have a time-dependent electric current, we consider the following ansatz for the gauge field \cite{Hashimoto:2013mua},
\begin{eqnarray}
\label{Axansatz}
A_x=-\int^tE(s)ds+h(t,z) \, ,\nonumber \\
E(t)=\frac{E_0}{2}\left(1+\tanh(\frac{t}{k})\right),
\label{Et}
\end{eqnarray}
where $E(t)$ a time-dependent electric field in the $x$ direction that increases from zero at $t_0$ to its final value $E_0$. The rate of energy injection to the system is controlled by the characteristic time scale $k$. Using the ansatz of eq.~(\ref{Axansatz}), the equation of motion (\ref{eomax}) would be a second-order nonlinear equation in both $t$ and $z$ for $h(t,z)$. Near boundary expansion of the equation of motion for $A_x$, leads to $h(z)\simeq b+a z^2+O(z^4)$. According to the AdS/CFT dictionary, the sub-leading term $a$ is proportional to the expectation value of the dual operator which translates to the current density of the boundary theory (the leading source term $b$ is a constant that sets to zero) \cite{Karch:2007pd}. Therefore, the time-dependent current in the boundary theory is proportional to the second derivative of $h(t,z)$ with respect to $z$ at the boundary,
\begin{eqnarray}
j(t) \propto \partial_z^2 h(t,z=0).
\label{jt}
\end{eqnarray}
In order to obtain the time-dependent electric current, we use the boundary conditions $h(t,z=0) = \partial_z h(t,z=0) = 0$ and initial conditions $h(t_0,z) = \partial_z h(t_0,z) = 0$ to solve the equation of motion numerically for different parameters such as the final value of the electric field, chemical potential, temperature, and the characteristic time scale.

In the following calculations, $T$ is the temperature of the initial thermal state given by the Hawking temperature eq.~(\ref{Htemp}), and we have set the AdS radius equal to 1 for numerical calculations. Also, $\mu_c=0.673$ and $T_c=0.27$ (in $GeV$ units) represent the critical chemical potential and critical temperature.
\begin{figure}
\center
\includegraphics[width=.7\linewidth]{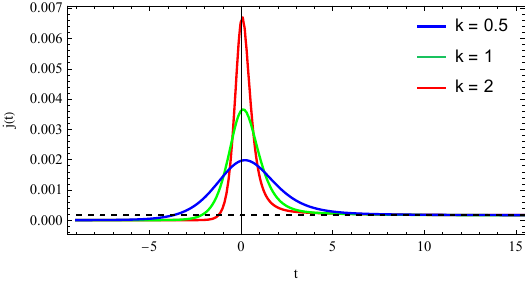}
\caption{\small (Color online) The time-dependent current $j(t)$ versus $t$ for $2\pi\alpha' E_0=0.001$ and $T=0.03\text{GeV}$ and $\mu=\mu_c$. Blue, green, and red curves correspond to $k=0.5,\,1, \text{and}\,2$ respectively. The dashed line is the static current $j_0$.}
\label{jtet}
\end{figure}

In Fig.~\ref{jtet} the time-dependent currents are shown for three different values of the characteristic time scale $k$. This figure is depicted for the lowest temperature of the holographic QCD model $T= 0.03~\text{GeV}$ associated with $\mu=\mu_c$ and $2\pi\alpha' E_0=0.001$. One can find that each current flow increases from zero to a maximum value and after a while, relaxes to the black dashed line $j_0$ defined by eq.~(\ref{js}). As mentioned before, the characteristic time scale $k$ defines the energy injection rate, and increasing $k$ leads to higher peaks in the current flow and lower equilibration times.

\begin{figure}[t!]
\center
\includegraphics[width=0.6\linewidth]{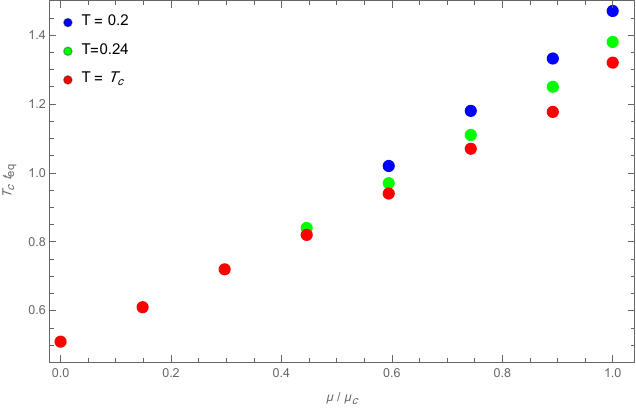}
\caption{\small The rescaled equilibration time in terms of $\mu/\mu_c$ for $2\pi\alpha' E_0=0.001$. Here, $k=0.25$ and blue, green and red lines correspond to $T=0.2,\,0.24, \text{and}\,0.27~\text{GeV}$ respectively.}
\label{tmu}
\end{figure}

The rescaled equilibration time $T_c t_{eq}$ versus the chemical potential ratio $\mu/\mu_c$ is plotted in Fig.~\ref{tmu} for $k=0.25$ and three different values of temperature. From this figure, one could observe that increasing chemical potential leads to increasing the rescaled equilibration time, and also, higher temperatures give rise to lower equilibration time.

Next, we are going to analyze the thermal quench in two different regimes of the characteristic time scale $k$ during which the electric field changes from zero to a constant value. The fast quenches with $k<1$ are more rapid quenches and result in faster energy injection and equilibration, while the slow quench is the opposite limit with large $k$ \cite{Buchel:2014gta}-\cite{Das:2014hqa}.

\begin{figure}
\begin{subfigure}{.5\textwidth}
\centering
\includegraphics[width=3in,height=2.3in]{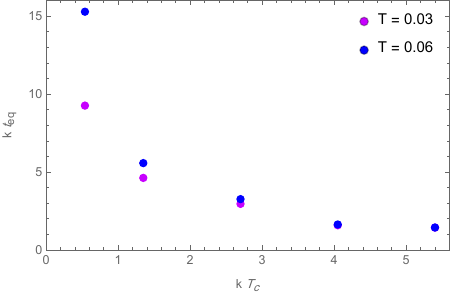}
\caption{}
\label{slow}
\end{subfigure}
\begin{subfigure}{.5\textwidth}
\centering
\includegraphics[width=3in,height=2.3in]{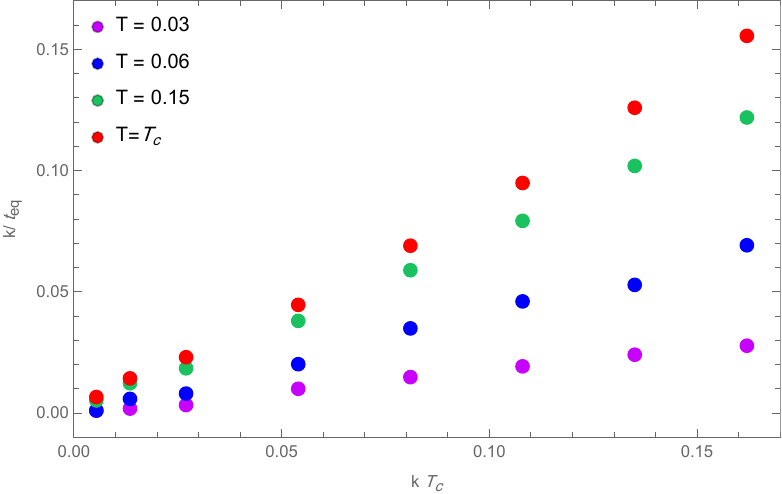}
\caption{}
\label{fast}
\end{subfigure}
\caption{(Color online) Slow and fast quenches for $2\pi\alpha' E_0=0.001$. (\subref{slow}) Slow quench regime: $k^{-1}t_{eq}$ with respect to $k$ for $T=0.03$ and $0.06$ (in $GeV$ units). (\subref{fast}) Fast quench regime: $k\,t_{eq}^{-1}$ with respect to $k$ for $T=0.03, 0.06, 0.15$, and $T_c$ (in $GeV$ units). }
\label{fastslow}
\end{figure}
The equilibration time for the slow quench regime is displayed in Fig.~\ref{slow} for $2\pi\alpha' E_0=0.001$. In this case, the electric field changes slowly over a long period, leading to an adiabatic response as the system has enough time to adjust to the energy injection and relax. From this figure, could find that for smaller values of $k$, the points corresponding to different temperatures are widely separated, while increasing the characteristic time scale leads to decreasing the rescaled equilibration time, and in large $k$ limit, the equilibration time scales as $t_{eq}\sim 1/k^2$ and the curves of different temperatures coincide. This could be regarded as a consequence of the zero entropy production in the adiabatic limit, analog to the slow quench regime behavior found in \cite{Buchel:2014gta}.

In Fig.~\ref{fast}, the rescaled equilibration time $k\,t_{eq}^{-1}$ is plotted with respect to $k$ for $2\pi\alpha' E_0=0.001$ at different temperatures. One could observe that decreasing $k$ leads to a universal behavior independent of temperature in which all data converge to a single point at the $k\rightarrow 0$ limit. Also, for small $k$, the slope of the line connecting the points of each $T$ is a constant proportional to the inverse of the temperature, indicating that in the fast quench regime, $t_{eq}\sim 1/T$ and increasing temperature leads to decreasing the equilibration time. It is consistent with the universal scaling behavior found in \cite{Buchel:2014gta}, where the relaxation time is set by the thermal time scale for fast quenches, regardless of the quenching rate. 

\begin{figure}
\centering
\includegraphics[width=0.6\linewidth]{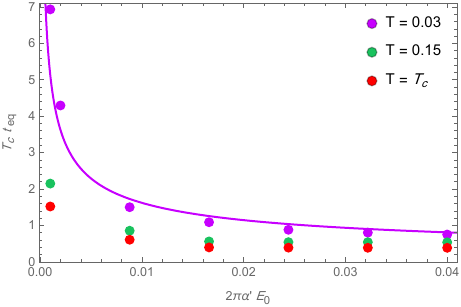}
\caption{\small (Color online) The rescaled equilibration time versus $2\pi\alpha'E_0$ for $k=0.25$ and $\mu=\mu_c$. The purple, green, and red points represent data for $T=0.03,\,0.15,\text{and}\, T_c$, respectively. The purple curve is a polynomial fit to $T=0.03$ data.}
\label{teqE}
\end{figure}

The rescaled equilibration time $T_c t_{eq}$ in terms of the dimensionless parameter $2\pi\alpha'E_0$ is plotted in Fig.~\ref{teqE} for $\mu=\mu_c$ and $k=0.25$ and different temperatures. The purple curve is a polynomial fitting to the lowest temperature data and has the following form,
\begin{eqnarray}
t_{eq} \simeq 0.6\,(2\pi \alpha' E_0)^{-1/2},
\label{teqlowT}
\end{eqnarray}

From this figure, one can observe that for larger electric field values, the equilibration time is less sensitive to temperature, and all curves approach the purple curve. Also, from Figs.~\ref{teqE} and \ref{tmu}, one could find that the equilibration time decreases by increasing temperature as well as decreasing chemical potential, and therefore, eq.~(\ref{teqlowT}) could be regarded as the upper bound of the equilibration time of the holographic QCD mode. One could compare eq.~(\ref{teqlowT}) with the thermalization time obtained in the Hashimoto-Oka model (in units of $2\pi \alpha^{\prime}$ and find the value $t_{eq}\sim 0.4\,\text{fm/c}$ for the equilibration time in the presence of an external electric field of order $10^4 MeV^2$\footnote{Strong electromagnetic fields expected to be produced in heavy-ion collisions \cite{Kharzeev:2007jp} with the magnitude of order $10^4\, MeV^2$ (see for example, \cite{Voronyuk:2011jd})}. This value is comparable with the isotropization time of the far-from-equilibrium non-isotropic plasma studied in \cite{Chesler:2008hg}, despite the different geometries, sources, and computational methods. In \cite{Chesler:2008hg}, a time-dependent shear (similar to $E(t)$ in eq.~(\ref{Et})) is considered in the geometry and the isotropization time is considered as the time when the transverse and longitudinal pressures deviate from their final values by less than $10\%$ and for transition times less than or equal to $2$ the isotropization time is estimated to be $0.5~\text{fm/c}$ for $T=0.35\text{GeV}$.

\section{Conclusion}
\label{sec5}

In this paper, we have investigated the equilibration  time of a massless quark in the presence of an external electric field by using the gauge/gravity duality. Our results demonstrate that for a strong electric field $E_0=10^4\text{MeV}^2$, the equilibration time would be about $0.4~\text{fm/c}$ which is comparable to the isotropization time obtained in \cite{Chesler:2008hg} despite different holographic models, sources, and implemented methods. This could be regarded as the characteristics of thermal quenches for the strongly coupled gauge theories reported in the literature for the fast quenches where the equilibration times for all temperatures converge to a single value for abrupt quenches.



\end{document}